\documentclass{SciPost}

\hypersetup{
    colorlinks,
    linkcolor={red!50!black},
    citecolor={blue!50!black},
    urlcolor={blue!80!black}
}

\usepackage[bitstream-charter]{mathdesign}
\usepackage{lineno}

\DeclareSymbolFont{usualmathcal}{OMS}{cmsy}{m}{n}
\DeclareSymbolFontAlphabet{\mathcal}{usualmathcal}

\fancypagestyle{SPstyle}{
\fancyhf{}
\lhead{\colorbox{scipostblue}{\bf \color{white} ~SciPost Physics Community Reports }}
\rhead{{\bf \color{scipostdeepblue} ~Submission }}

\fancyfoot[C]{\textbf{\thepage}}
}

\usepackage{color}
\definecolor{mit-red}{rgb}{0.64,.12,0.2}
\definecolor{darkred}{rgb}{1.0,0.1,0.1}
\definecolor{darkgreen}{rgb}{0.1,0.7,0.1}
\definecolor{darkblue}{rgb}{0.1,0.1,1.0}

\DeclareRobustCommand{\Sec}[1]{Sec.~\ref{sec:#1}}
\DeclareRobustCommand{\Secs}[2]{Secs.~\ref{sec:#1} and \ref{sec:#2}}

\DeclareRobustCommand{\Fig}[1]{Fig.~\ref{fig:#1}}

\DeclareRobustCommand{\Eq}[1]{Eq.~(\ref{eq:#1})}

\DeclareRobustCommand{\Reference}[1]{Ref.~\cite{#1}}

\DeclareRobustCommand{\Def}[1]{Def.~\ref{def:#1}}

\usepackage{titling,multicol}
\usepackage{tcolorbox}
\tcbuselibrary{theorems}
\newtcbtheorem[number within=section]{mydef}{Definition}%
{colback=red!5,colframe=red!35!black,fonttitle=\bfseries }{def}

\begin{document}

\pagestyle{SPstyle}

~\vspace{-0.45in}

\hfill {\footnotesize MIT-CTP/6058}

\vspace{0.15in}

\begin{center}{\Large \textbf{\color{scipostdeepblue}{
%%%%%%%%%% TODO: Write your article's title here
Interpreting ``Interpretability'' and Explaining ``Explainability'' in Machine Learning in Physics\\
}}}\end{center}

\begin{center}
Part of the VERaiPHY Initiative
\end{center}

\begin{center}\textbf{
% Use (full) first name (+ middle name initials) + surname format.
% Separate subsequent authors by a comma, omit comma and use "and" for the last author.
% Mark the corresponding author(s) with a superscript symbol in this order
% \star, \dagger, \ddagger, \circ, \S, \P, \parallel, ...
Rikab Gambhir\textsuperscript{1,$\star$},
Luisa Lucie-Smith\textsuperscript{2,$\star$}, and
Jesse Thaler\textsuperscript{3,4,5,6,$\dagger$}
}\end{center}

\begin{center}
{\bf 1} Department of Physics, University of Cincinnati, Cincinnati, Ohio 45221, USA
\\
{\bf 2} Hamburger Sternwarte, Universit{\"a}t Hamburg, Gojenbergsweg 112, 21029 Hamburg, Germany
\\
{\bf 3} Center for Theoretical Physics -- a Leinweber Institute, Massachusetts Institute of Technology,  Cambridge, MA 02139, USA
\\
{\bf 4} Institut des Hautes \'Etudes Scientifiques, 91440 Bures-sur-Yvette, France
\\
{\bf 5} Institut de Physique Th\'eorique, CEA Paris-Saclay, 91191 Gif-sur-Yvette, France
\\
{\bf 6} The NSF Institute for Artificial Intelligence and Fundamental Interactions, USA
%%%%%%%%%% END TODO: AFFILIATIONS
% Provide email address of corresponding author(s)
\\[\baselineskip]
$\star$ Lead authors %\href{mailto:email1}{\small email1}
\,\quad
$\dagger$ Advisor %\href{mailto:email2}{\small email2}
\end{center}

\section*{\color{scipostdeepblue}{Abstract}}
\textbf{\boldmath{%
We review the concepts of interpretability and explainability as they apply to machine learning in physics.
We define \emph{interpretability} as concerning the structural transparency of a model (the ability to understand or approximate its inner workings) and \emph{explainability} as concerning the scientific content of a model (the ability to map it onto domain knowledge).
We discuss the trade-offs each entails (interpretability vs.\ expressivity; explainability vs.\ adaptability), the contexts in which each is needed, and the intrinsic and post-hoc tools available for achieving them.
Throughout, we emphasize that machine-learned models are subject to the same scientific questions as classical models, differing only in scale, and that interpretability and explainability are best understood as deliberate modeling choices rather than inherent properties.
We also emphasize the importance of task specification and intervention plans as a core aspect of model design.
}}

\vspace{\baselineskip}

% %%%%%%%%%% BLOCK: Copyright information
% % This block will be filled during the proof stage, and finilized just before publication.
% % It exists here only as a placeholder, and should not be modified by authors.
% \noindent\textcolor{white!90!black}{%
% \fbox{\parbox{0.975\linewidth}{%
% \textcolor{white!40!black}{\begin{tabular}{lr}%
%   \begin{minipage}{0.6\textwidth}%
%     {\small Copyright attribution to authors. \newline
%     This work is a submission to SciPost Phys. Comm. Rep. \newline
%     License information to appear upon publication. \newline
%     Publication information to appear upon publication.}
%   \end{minipage} & \begin{minipage}{0.4\textwidth}
%     {\small Received Date \newline Accepted Date \newline Published Date}%
%   \end{minipage}
% \end{tabular}}
% }}
% }
% %%%%%%%%%% BLOCK: Copyright information

% %%%%%%%%%% TODO: LINENO
% % For convenience during refereeing we turn on line numbers:
% \linenumbers
% % You should run LaTeX twice in order for the line numbers to appear.
% %%%%%%%%%% END TODO: LINENO

% #####################################################
% ########## Paragraphs in TOC. Delete later! #########
% #####################################################

\setcounter{tocdepth}{2}

\vspace{10pt}
\noindent\rule{\textwidth}{1pt}
\tableofcontents
\noindent\rule{\textwidth}{1pt}
\vspace{10pt}

% Write your article contents here, starting from first \section.
% An example structure is given below.

\section{Introduction}

The success of highly expressive machine learning (ML) models in the physical sciences, such as deep neural networks and more recently large language models (LLMs), has raised fundamental questions about their interpretability, their scientific reliability, and their role in scientific progress.
This discussion relates to a broader topic of interest within the artificial intelligence (AI) community: how to define, evaluate, and ensure desirable properties of increasingly complex and opaque models. 
The physics community, especially in the context of ML, often interchangeably makes use of words such as \textit{interpretability}, \textit{explainability}, \textit{trustworthiness}, \textit{transparency}, \textit{fairness}, \textit{safety}, and
\textit{understandability}~\cite{interp_review}. 
All these terms point to difficult-to-pin-down properties that we would like ML models to satisfy, yet which they do not typically possess out of the box, unlike our ``classical'' scientific models.\footnote{By ``classical'', we mean models created by scientists without using ML techniques.}

In this review, our goal is to pin down what exactly these properties are, with a focus on \emph{interpretability} and \emph{explainability} and their role in both ML and classical models for high energy physics (HEP), astrophysics, and cosmology.
In particular, we make a distinction between interpretability and explainability, following the definitions introduced in \cite{Lucie-Smith:2022uvv}.\footnote{Related terms that are often used are \emph{intelligibility} and \emph{understandability}. For the purposes of this review and for simplicity, we will consider interpretability and explainability to be the primary axes of interest.}
Fundamentally, \textbf{\textit{interpretability is about a model's structure}}, whereas \textbf{\textit{explainability is about a model's scientific content}}.
These two desiderata -- interpretability and explainability -- are characteristics of ML models that fall under the umbrella of explainable AI (XAI). 
XAI has gained much attention in fields of the physical sciences and beyond, where being able to explain the decision-making process is a non-negotiable prerequisite. 
These include applications of AI to fundamental science, healthcare, crime science, neuroscience, and legal and ethical policies.

Our fundamental axiom will be that \emph{ML models are no different than classical models, except in size}.\footnote{In saying this, we are side-stepping the interesting world of neural scaling laws, when the number of parameters in a model is so large that the model itself is its own physical system.}
Certainly, classical models like Newton's Laws, the Standard Model (SM) of particle physics, or General Relativity \emph{feel} different than ML models, owing to the fact that these models are determined by physical principles, even though they may still contain parameters that need to be calibrated from data.
We will argue that this difference can be rephrased in terms of interpretability and explainability of a given model.
With this identification, we demonstrate that issues of interpretability and explainability are \emph{not} an idiosyncrasy of modern ML techniques, but rather broad overarching questions to ask about model selection and design across all physical sciences.

In our discussion, we emphasize the importance of \emph{task specification} and \emph{intervention planning} in model design. 
Interpretability and explainability are not always necessary, and this determination is highly problem dependent and subjective.
Since interpretability and explainability are user-defined concepts, we emphasize that one can design an ML model, loss function, or training technique in a way that aligns with their desired scientific
goals, which may or may not include interpretability and explainability.
That said, we avoid making prescriptive statements about how to specifically achieve interpretability or explainability, due to their problem-dependent nature.

The rest of this review is organized as follows.
In \Sec{definitions}, we provide definitions for interpretability and explainability, and we discuss their trade-offs.
In \Sec{interpretability} and \Sec{explainability}, we focus on the concepts of interpretability and explainability in greater detail, respectively.
We discuss examples of models that achieve these properties, the trade-offs required to obtain them, and the contexts in which they are most needed. 
In \Sec{intrinsic_interpretability}, we discuss intrinsic and post-hoc methods for achieving interpretability and explainability.
In \Sec{task}, we discuss task specification and intervention planning as prerequisites for any interpretability or explainability effort.
We conclude in \Sec{conclusion}.

This article contributes to VERaiPHY (Validation \& Evaluation for Robust AI in PHYsics), a PHYSTAT review series establishing verification and validation standards for ML across particle physics, astrophysics, and cosmology.

% ##########

\section{Foundations of Interpretability and Explainability}
\label{sec:definitions}

In this section, we define interpretability, explainability, and related concepts. 
These definitions align with their common usage in the AI community. 
Whether or not these are \emph{good} definitions remains an open question in the field, but for the purpose of this review, we will work with these established definitions while recognizing that the community's understanding of these concepts continues to evolve.

\subsection{Core Definitions}

\paragraph{Definition of a \textit{model} |}
For our purposes, a model is any fixed function $f_\theta$ that produces a prediction $f_\theta(x)$ from an input measurement $x$ given parameters $\theta$.
Here, the model parameters $\theta$ are fixed by past or auxiliary measurements of $x$.
The procedure by which these are fixed, called ``training'' or ``fitting'', is itself part of the model specification.
Models can be either deterministic or probabilistic, and indeed most models relate in some way to the probability of measuring $x$ given parameters $\theta$.
The model $f$ can be written explicitly or implicitly as the solution to some mathematical procedure or numerical simulation.
The latter is incredibly common in HEP, astrophysics, and cosmology.
In HEP, state-of-the-art predictive models involve a pipeline of event generators and simulations, including \textsc{MadGraph}~\cite{Alwall:2011uj}, \textsc{Pythia}~\cite{Bierlich:2022pfr}, and \textsc{Geant}~\cite{GEANT4:2002zbu} for various components of a collider analysis.
In astrophysics and cosmology, the theoretical predictions of cosmological probes of the early- and late-time Universe are obtained through Boltzmann solvers \cite{Lewis:1999bs, 2011JCAP...07..034B}, as well as gravitational and hydrodynamical simulations of the evolution of the Universe over cosmic times \cite{Springel:2005mi, 2014MNRAS.444.1518V, 2015MNRAS.446..521S, 2019MNRAS.490.3196P}.
All of these models are based on a combination of first-principles physics, phenomenological modeling, and free-parameter fitting. 
In this context, an ML model can be simply viewed as a very flexible model $f_\theta$ with a large number of parameters $\theta$, whose functional form is broad enough to universally approximate large classes of functions. 
From this perspective, there is no fundamental distinction between ML models and ``classical'' ones: both define a mapping $f$, differing only in size, specific functional forms, and practical implementation details.

\paragraph{Definitions of \textit{interpretability} and \textit{explainability} |} 
Interpretability and explainability concern \emph{meta questions} about models. 
Rather than asking how \emph{good} the model is (as measured by predictive power, error rate, or some other empirical measure), we are concerned with asking how well we can understand the model, or how it relates to knowledge drawn from previous models and data.
We define \emph{interpretability} and \emph{explainability} as follows~\cite{Lucie-Smith:2022uvv}:

\hspace{1cm}
\begin{mydef}{Interpretability}{interpretability}
  \textbf{Interpretability} concerns the ability to understand, or approximate, the inner workings of a model and how it reaches its output.\\

\emph{``Given a model $f_\theta(x)$, can we read off what $f$, or parts of $f$, represent? Can we approximately characterize how the model outputs relate to the model inputs and/or parameters?''}\\

This concerns the relationship between model inputs, parameters, and outputs, and the training procedure.\\

Keywords: \textit{structure}, \textit{mechanistic}, \textit{computation}, \textit{approximability}, \textit{form}.
\end{mydef}

\hspace{1cm}

\begin{mydef}{Explainability}{explainability}
  \textbf{Explainability} concerns the ability to map our interpretation of a model onto existing knowledge in the relevant scientific domain.\\

\emph{``Given a model $f_\theta(x)$ that we have interpreted, can we associate the learned $\theta$, or functions of $\theta$, with domain knowledge?''}\\

Here, \emph{domain knowledge} refers to prior information about the underlying physical system at hand, often acquired from another model or experimental/observational data.\\

Keywords: \textit{domain knowledge}, \textit{scientific content}, \textit{information}, \textit{context}, \textit{semantics}.
\end{mydef}
\hspace{1cm}

We will unpack these definitions in significant detail in \Secs{interpretability}{explainability}, and devote the rest of this section to further establishing the taxonomy of interpretability and explainability.
Interpretability and explainability are highly interlinked and correlated concepts.
In many cases, the goal of interpretability is explainability, as the latter cannot be achieved without the former. 
Additionally, both properties are achievable via a vast spectra of techniques and model architectures, with many tools falling under both umbrellas of interpretability and explainability depending on their use case.

\paragraph{Definitions of  \emph{intrinsic} and \emph{post-hoc} |}
Interpretability and explainability can each be subdivided into two flavors -- \emph{intrinsic} and \emph{post-hoc} -- based on how they are applied to models. 
These two categories encapsulate the principal methodologies for achieving interpretability and explainability in models: either by design or by analysis after training.
We define \emph{intrinsic} and \emph{post-hoc} as follows:

\hspace{1cm}
\begin{mydef}{Intrinsic}{Intrinsic}
  \textbf{Intrinsic} interpretability or explainability concerns the design of the model $f_\theta(x)$.\\

  \emph{``Can the model $f_\theta(x)$ be constructed to be interpretable or explainable for a given application by embedding structural constraints or domain knowledge directly into $f_\theta$?''}
\end{mydef}
\hspace{1cm}

\begin{mydef}{Post-Hoc}{posthoc}
  \textbf{Post-hoc} interpretability or explainability concerns the use of analysis techniques applied to an existing model $f_\theta(x)$.\\

  \emph{``Given a model $f_\theta(x)$, can we interpret its behavior or explain its results a posteriori? For example, can we use additional metrics to distill, approximate, or explain the model in relation to its science domain?
  }
%
%\Rikab{TODO: Be consistent about I vs We}
\end{mydef}
\hspace{1cm}

We will discuss these methods further in \Sec{intrinsic_interpretability}. 
We note here that it is often the case that both intrinsic and post-hoc methods usually only achieve interpretability directly, and  explainability is only achieved through having access to interpretable structures.

\subsection{The Landscape of Interpretability and Explainability}
While interpretability and explainability often go hand-in-hand, models can span the entire interpretability-explainability plane.
Exploring each quadrant of thisplane,\footnote{We emphasize that interpretability and explainability are not \emph{orthogonal} axes on this plane. Think of them as correlated curvilinear coordinates.} we have models that are:

\paragraph{Interpretable and explainable |} ``Classical'' physics models are the quintessential example of interpretable and explainable models. 
Models like Newtonian gravity, the Standard Model, or General Relativity are (relatively) easy to compute and extract scientific information from, as they are derived from first principles.
Simple models derived from traditional ML techniques, such as shallow decision trees or simple linear models may also fall into this category.
Here, parameters are semantically meaningful and are easy to relate to outputs.

\paragraph{Interpretable but not explainable |} While interpretability focuses on understanding model structure or behavior, explainability is typically harder to achieve but a more meaningful goal, as connecting model behavior to underlying physical principles or causal mechanisms requires inputs beyond the model's mechanistic structure.
Thus, interpretability without explainability is relatively common.
One can imagine various ways to interpret the features learned by a convolutional layer, e.g., by inspecting the activations, the learned kernels, or their gradients. 
However, such inspections often remain qualitative and visual in nature, making it difficult to relate them to physically meaningful features in the relevant scientific domain. 

\paragraph{Explainable but not interpretable |} In general, explainable-but-not-interpretable models do not exist. This is because explainability comes as a second step after interpretability: one first identifies a relevant component or behavior of the model (interpretability) and then relates this to domain knowledge (explainability). In other words, some degree of interpretability is always required before a model can be explained.

For example, a QCD prediction at 10-loops is difficult to interpret because it involves a series of nigh-impossible integrals. Yet behaviors like $\alpha_s$ scaling or locations of poles or logarithms can still be identified, and this partial interpretation is what makes the model explainable in terms of its underlying theory. A flow-based model surrogate offers another example: the mechanistic inner workings of the flow remain opaque, but some interpretability can still be gained empirically -- for instance, by probing the model's response to different noise levels or examining its behavior across various limits and regimes. Such a surrogate is explainable so long as its range of validity and failure modes are understood in terms of the underlying physics it is trying to mimic.

\paragraph{Neither interpretable nor explainable |} Models that are neither interpretable nor explainable have a questionable role in science.\footnote{We make a bolder conjecture: If a model appears to have scientific or predictive value, then there \emph{must} be something explainable and/or partially interpretable about it.}
These models include proprietary black boxes such as common LLMs, which can neither be interpreted nor explained since the user does not have access to them.
These models are often better used as tools or assistants to science rather than as scientific models in and of themselves.
Non-physics examples include models like the celebrated AlphaFold~\cite{Jumper2021}, which is extremely difficult to interpret, but still makes predictions anchored to biochemical principles.\footnote{Specifically, AlphaFold does not just predict a 3D protein geometry directly from raw sequences, but rather uses a built-in inductive bias to encode the fact that parts of the protein physically close together in 3D space have correlated mutation histories. This connection to the protein's evolutionary mutation history is part of the domain knowledge used to explain AlphaFold's predictions. This does not contradict our definition of explainability requiring interpretability, as the fact that we can discuss these mechanisms at all required interpreting AlphaFold's structure.} 

\begin{figure}
    \centering
    \includegraphics[width=0.7\linewidth]{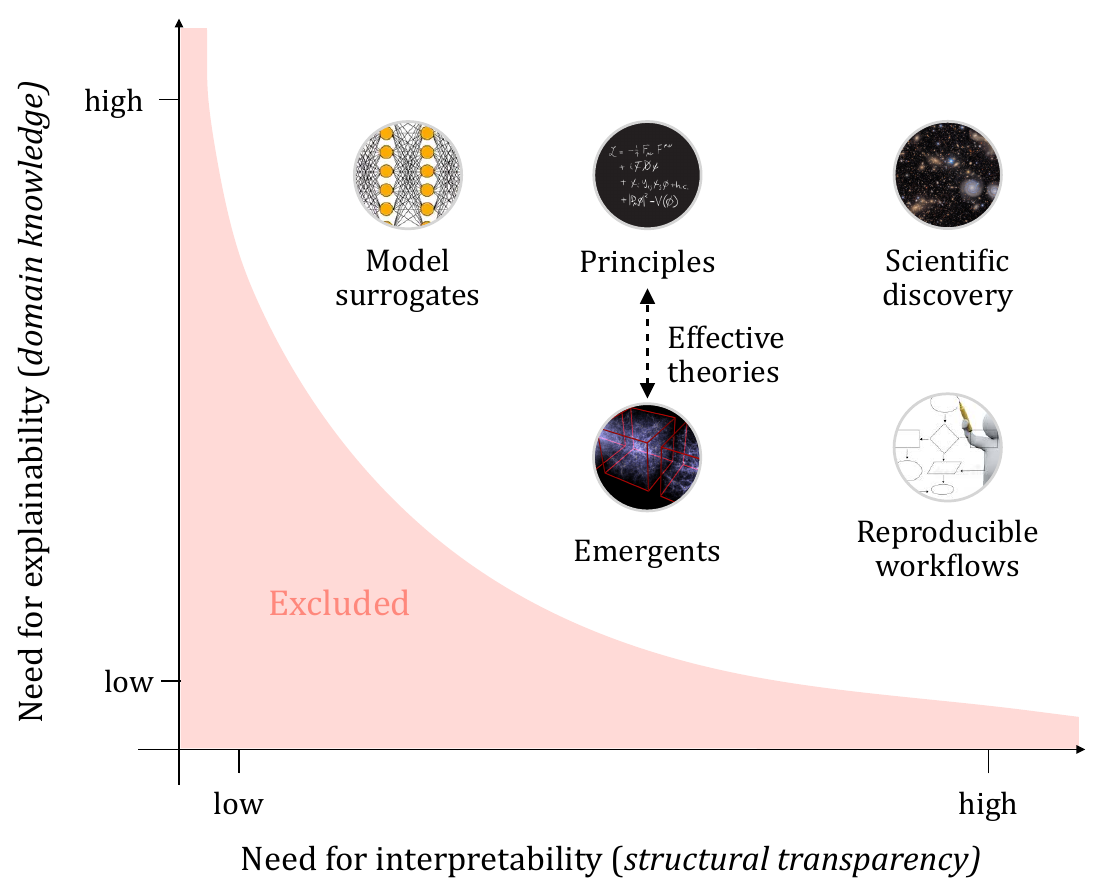}
    \caption{
    Examples of scientific goals and existing models in HEP, astrophysics, and cosmology plotted by their need for interpretability versus explainability. The bottom-left region of the plot is ``excluded'', as all scientific goals require some degree of interpretability and explainability.}
    \label{fig:graphics_exp_vs_int}
\end{figure}

\subsection{Scientific Goals Requiring Interpretability and/or Explainability}
For which applications are interpretability, explainability, or both truly required?
In \Fig{graphics_exp_vs_int}, we show examples of scientific goals in HEP, astrophysics, and cosmology based on their relative need for interpretability and explainability.
In our view, there are no scientific applications for which both interpretability and explainability are unnecessary (bottom-left region of \Fig{graphics_exp_vs_int}). 
In scientific research, models that cannot be (even partially) understood mechanistically, approximated, explained, validated, or inquired using prior domain knowledge or prior models, are of little-to-no scientific value.
Exploring each of the scientific goals outlined in \Fig{graphics_exp_vs_int}, we have:

 \paragraph{Scientific discovery | }A genuine scientific discovery must be both highly interpretable and explainable.
As we will explore further in \Secs{interpretability}{explainability}, interpretability is necessary to check and trust the result, and explainability is necessary to declare the result a discovery in comparison to previous scientific knowledge. 
Without interpretability, a claimed scientific discovery is nebulous and vague, and without explainability, it is just a contextless list of numbers, fits, and measurements.

\paragraph{Principles and emergents |} Explainability is an important aspect of highly complex models where the goal is to implement first-principles physics.
In many cases, first-principles models are inherently complex and difficult to interpret. 
For instance, lattice gauge theory, as a systematic implementation of quantum field theory, is and must be highly explainable, even if it is numerically highly complex. 
For the same reason, a detailed cosmological simulation based on first-principles Newtonian gravity is also fully explainable.
However, even with first-principles and fully detailed microphysics, it often remains difficult to explain \emph{macroscopic} or \emph{emergent} phenomena, such as the formation of hadrons or the specific shape or morphological features of galaxies. 
That is, there is a ``hierarchy'' of explanations one might be interested in for different types of scientific questions.
The first-principles physics is not enough to explain these collective phenomena, and often effective theories or phenomenological ``second-principles'' models are needed to bridge the explainability gap.

\paragraph{Model surrogates |}  \emph{Acceleration} and \emph{automation} are common tasks for ML in science.
Acceleration typically involves making a pre-existing model computationally faster~\cite{SpurioMancini2022, Alsing2020, Lehman_2026, Arico_2022}.
These pre-existing models are often either first-principles or phenomenological models rooted within established scientific domain knowledge, or instrumental or detector models of systematics such as calibrating the point spread function of a telescope or unfolding detector distortions in collider data. 
The ML model then acts as a ``surrogate'' of the original model to accelerate computation. 
In this case, interpretability is not needed to a high degree: a complex ``black-box model'' can be used, provided its predictions accurately reproduce those of the original model while requiring only a fraction of the compute resources. 
On the other hand, explainability is crucial: this is either already provided to a large extent by the original model which the surrogate is trained to reproduce, or can be achieved by thoughtful validation/evaluation of the model in different limiting regimes to establish the range of validity and limitations of the surrogate.

\paragraph{Reproducible workflows | } Replicability in science is a goal for which interpretability is paramount.
Regardless of the scientific correctness of the procedure, it is essential that workflows are reproducible, which requires a degree of interpretability for scientists to be able to understand each other's work. 
Interpretability also lends itself to intervention. 
An interpretable pipeline, such as a detector simulation or an analysis framework, is one that is easy to diagnose and debug if problems should arise.

\section{Interpretability}
\label{sec:interpretability}

Interpretability concerns the ability to understand the \textit{structural and mechanistic} inner workings of ML models and how the models reach their final predictions. 
It is a mechanistic statement about a user's ability to compute, approximate, or describe the model, and can be viewed as analogous to the ``description length'' (i.e.\ Kolmogorov complexity~\cite{kolmogorov1965three}) of a model.
\emph{In general, a high degree of interpretability is what separates classical models from (more complex) ML models, and the assignment of semantic meaning to model components is part of how one approximates behavior in interpretable models}.

It is important to note that interpretability, as defined in \Def{interpretability}, is entirely data and domain-knowledge agnostic.
Interpretability is primarily a feature of the \emph{form} of the model $f$ and its dependence on $\theta$ and $x$, not of the specific values of $\theta$ or $f_\theta(x)$.
That is, Newton's Law is equally interpretable whether $G_N = 6.67\times 10^{-11}$ N (m/kg)$^2$ or $1.0$ N (m/kg)$^2$. We say ``primarily'' because there exist cases where special parameter values (e.g.\ zero) can render a model more transparent or more simple than a generic value.\footnote{Many models have special parameter settings where they may simplify, e.g.\ both Newton's Law and dense neural networks become trivial if all their parameters are set to zero. This is still a domain-agnostic statement, however. \textit{``If the parameter were zero, the model structure would simplify or change''} is a statement about the model structure independent of data or the actual value of the parameter. The presence of special parameter values is a question of explainability.} 
This distinction is important, as the ability to easily compute or approximate a model well is orthogonal to the scientific content of the model, which falls under the umbrella of explainability.

A major aspect of our definition of interpretability is what XAI literature refers to as \emph{mechanistic interpretability}~\cite{olah2020zoomin, elhage2021mathematical, bereska2024mechanistic}, which concerns the ability to analyze a model's computations in terms of basic mathematical processes, algorithms, or circuits. 
However, we resist restricting our definition to solely mechanistic interpretability, as a full understanding of a model's structure also relies on \emph{access} (can the model be queried?), \emph{openness} (are parameters and code released?), \emph{documentation} (is the training procedure described? Can the model be easily described?), \emph{reproducibility} (could the model be retrained?), which are axes orthogonal to mechanistic interpretability. 
An inaccessible model cannot be interpreted, regardless of how mechanistically simple it is.
There are many ongoing efforts to improve the accessibility and transparency of scientific models, such as the promotion of FAIR (findable, accessible, interoperable, and reusable) data principles~\cite{Duarte:2022job}. 

In the rest of this section, we aim to explore the large spectrum of interpretability, discuss the tradeoff between interpretability and expressivity, and finally pose questions about when interpretability is needed, and when it is not.

\subsection{Spectrum of Interpretability Models and Tools} 

\begin{figure}
    \centering
    \includegraphics[width=\linewidth]{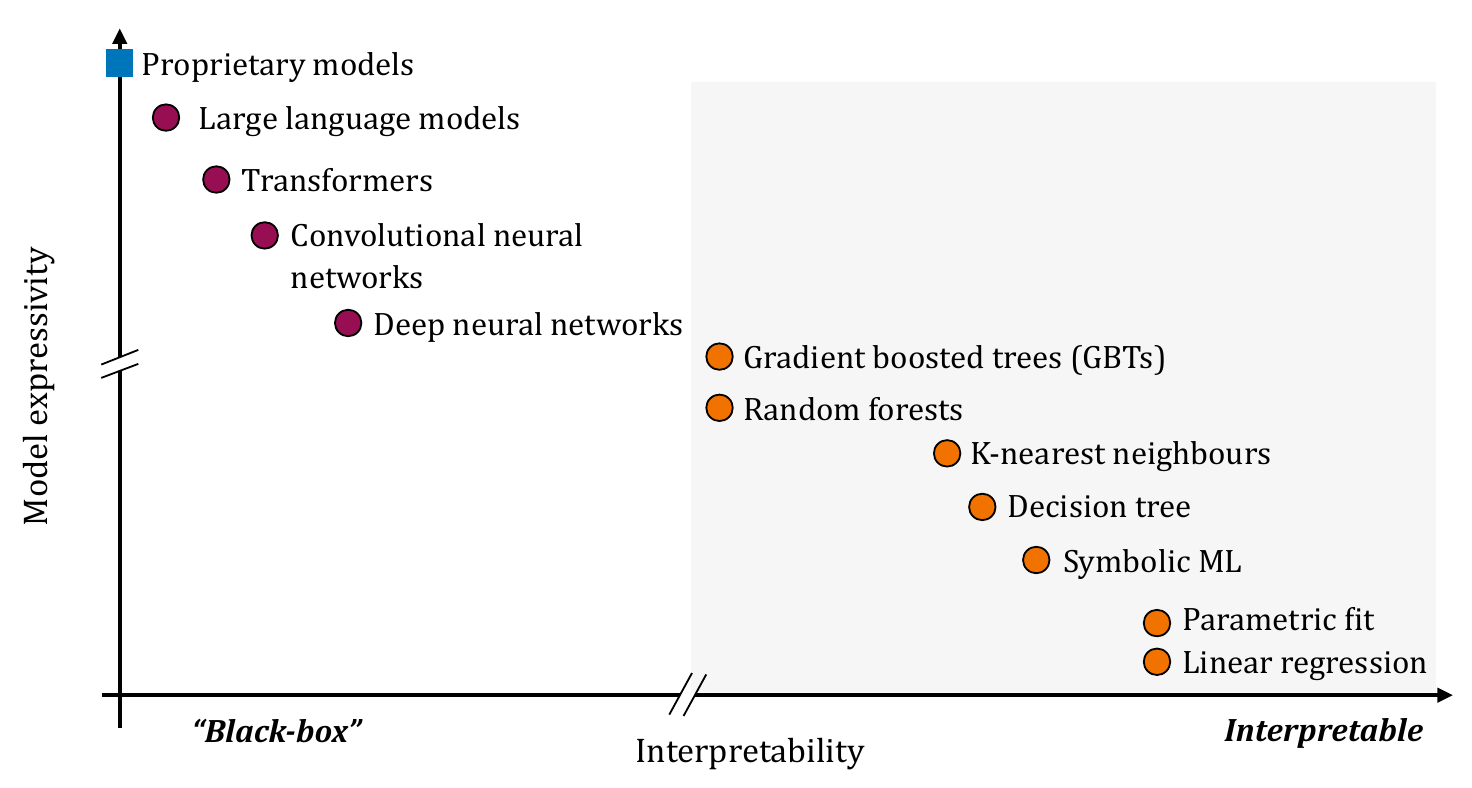}
    \caption{The trade-off between interpretability and expressivity. Models that are interpretable by-design (orange circles) have limited expressivity, as they are constrained to simple (e.g.\ linear or binary) relationships.  By contrast, highly parametrized models (purple circles) are extremely expressive and flexible, yet offer little insight into how and why they reach certain predictions, and the most extreme example (blue square) is completely uninterpretable.}
\label{fig:graphics_interpretability}
\end{figure}

We start by ranking the relative structural interpretability of several common models along the horizontal axis of \Fig{graphics_interpretability}, and later discuss the trade-off between interpretability and model expressivity (on the  vertical axis). 
We remind the reader that the definition of interpretability is subjective, as the model designer or user decides what they can interpret, and thus this ranking should not be taken as gospel.

\paragraph{Not interpretable: inaccessible models |} 
The extreme ``least interpretable'' models in the leftmost part of \Fig{graphics_interpretability} are proprietary models (blue square):
these are models for which there is an $f_\theta(x)$, but the end user does not have access to the function (even in code) or know how it was produced or fit. 
These models are less common in scientific applications of ML for physics, but serve as a useful extreme case to illustrate what non-interpretability looks like, and to illustrate the relation between interpretability and model trust (more details on this in \Sec{whentocareinterpretability}).
This is not to say that these models have no place in physics: indeed, proprietary LLMs have been used to construct agents or coding assistants for physics applications~\cite{Bakshi:2025fgx, Menzo:2025cim, Plehn:2026gxv}. Yet, the implications of their use in science are the subject of recent debate \cite{hogg2026astrophysics}.

\paragraph{Barely interpretable: deep learning |}Moving rightwards along the interpretability axis, we arrive at common deep learning models (purple circles in \Fig{graphics_interpretability}), spanning architectures such as transformers, convolutional and dense neural networks. 
These typically include a huge number (reaching even millions, or even billions in commercial tools) of parameters; this in turn makes it difficult to understand which features are being extracted from the input data and how these features are then combined into predictions. 
These models are often referred to as ``black boxes''.
Unlike proprietary models, these models typically exist as a deterministic computer code one can fully inspect, so one can \emph{technically} interpret them.
However, this is nigh-impossible in practice, analogous to how one could \emph{in principle} deterministically predict or approximate the output of a pseudo-random number generator, but \textit{in practice} cannot.\footnote{See \Reference{siu2023stlsurprisinglytrickylogic} for a study on humans' poor performance on interpreting formal specifications of algorithms.} 
An apparent exception is often claimed for transformers, where it is argued that model behavior can be understood by inspecting self-attention matrices, i.e.\ by analyzing which tokens attend to others and inferring structure or reasoning from these patterns. 
However, it remains unclear to what extent such interpretations faithfully reflect the underlying computations~\cite{jain2019attentionexplanation}, or if the power of transformers is due to simple nonlinear multiplication~\cite{Gambhir:2024dtf}.
For most deep learning applications, this is where post-hoc methods enter the spotlight, as one can instead aim to understand broad, approximate features of the model.
Post-hoc interpretability techniques will be discussed in more detail in \Sec{posthoc}.

\paragraph{Somewhat interpretable: simple ML |}
On the more interpretable side of the spectrum (rightmost region of \Fig{graphics_interpretability}) are more traditional, ``simpler'' ML models such as ensembles of decision trees and clustering algorithms.
We deem these as interpretable because their design makes it possible to understand how each input directly determines the output, for example via simple decision trees or linear transformations.

\paragraph{Very interpretable: parametric fits \& classical models | }Among the most interpretable models are low-dimensional parametric fits.
Not only are they extremely simple, but also one has the freedom to choose whichever functional form deemed to be interpretable for the problem at hand.
This category includes many classical physics models: for example, Newton's Law of Gravity is a simple power law with a single free parameter $G_N$.
These models are often so easy to compute or approximate that we can predict emergent, generalized behavior, or judge how each individual parameter will affect the output.
For example, Newton's Law is simple enough that one can predict closed orbits purely from its structure alone, regardless of the value of $G_N$ or access to any data. 
Similarly, since forces are directly proportional to this parameter $G_N$, we can assign $G_N$ a semantic meaning -- the ``strength of gravity'' -- so that we may easily approximate how the properties of gravitational systems change with this parameter without needing to do a detailed mechanistic computation.

\subsection{Trade-offs: Interpretability vs. Expressivity}\label{sec:interpretability_trade_off}

\paragraph{Model expressivity |}While intrinsically interpretable models -- those whose structure can be easily understood -- are often desirable, there is a fundamental ``cost'' associated to interpretable models in the form of expressivity.
A model's expressivity is defined as the ``volume'' of space of functions able to be reached: higher expressivity allows the model to fit more complicated patterns in the data, while lower expressivity restricts it to simpler patterns.
Model $A$ is more expressive than Model $B$ if at least every function approximated by Model $B$ can also be approximated by Model $A$.
A major aspect of the ML revolution is that neural networks, and all of their variants, are capable of approximating very large classes of functions efficiently, as given by a series of universal function approximation theorems.

Expressivity is desirable when the function one wishes to approximate is either unknown or highly mathematically complex.
For example, the precise structure of language is both unknown and highly complex, and thus highly expressive architectures such as transformers are ideal for flexibly capturing such complexity.
However, an unknown or highly complex model is, by definition, uninterpretable, leading to a trade-off between interpretability and expressivity.

\paragraph{The trade-off |}
In \Fig{graphics_interpretability}, we explore the trade-off between model expressivity and interpretability of the common ML models and techniques discussed above.
Less interpretable models, such as transformers, tend to be highly expressive, explaining their now ubiquitous use for a wide range of scientific problems.
On the other extreme, parametric fits tend to be highly interpretable (by design, because the user can choose them), but are highly restrictive in their functional forms.
Often, the best choice in a scientific context is the most interpretable model that is just expressive enough to capture the model features of interest. 
However, this is a highly domain- and problem-dependent task that fundamentally depends on whether or not interpretability is needed for that specific application, and is an important aspect of task specification.
For instance, overly expressive models, which might naively overfit, can be rescued if the model's inductive bias aligns with the data, but this is an issue of explainability, which we will discuss further in \Sec{care_explainability}.

\subsection{When Should One Care About Interpretability?} \label{sec:whentocareinterpretability}

\paragraph{When not to care: expressivity, surrogates, and benchmarks |}
The decision to address interpretability, either through intrinsic or post-hoc methods, is a modeling choice that must be carefully considered.
First, we argue that one cannot have \emph{zero} regard for interpretability; otherwise, one would be satisfied using a random unknown model handed to them by a stranger without context.
It is not science if one does not even know what their model \emph{is}.
However, some applications do not require a high degree of interpretability.
\begin{itemize}
\item \textbf{Expressivity}:  In cases where expressivity is important, interpretability can be deprioritized.
One cannot expect, for example, that either the SM or an ML proxy to it can be written in terms of simple elementary functions or simple latent spaces, even if we have large amounts of domain knowledge and full control over the physical parameters.
\item \textbf{Surrogates}:
Another example where interpretability is less important is when we already have a known function or process, and we simply want a fast or differentiable surrogate for it, as is often the case in simulation-based inference~\cite{Cranmer_2020}, faster simulation and data automation (e.g.\ the CaloGEN challenge~\cite{Krause:2024avx} and photometric redshift estimation \cite{CollisterLahav2004, Carrasco2013}), or faster theoretical computation (e.g.\ surrogate models in lattice gauge theory~\cite{Abbott:2025kvi, Abbott:2026ylv, Lawrence:2025rnk} or neural-based emulators for summary statistics in cosmology \cite{SpurioMancini2022,Arico_2021, 2020ApJ...901....5B}).
\item \textbf{Benchmarks}:
When one is solely motivated to optimize some performance metric or benchmark (e.g.\ classification performance), interpretability can be sacrificed.
Even in these cases, though, some small amount of interpretability is required; otherwise, one can often find trivial solutions that optimize the metric exactly (e.g.\ look-up tables), which is often not what one truly wants, which we will discuss further in \Sec{task}. 

\end{itemize}

\paragraph{When to care: trust, explainability, and efficiency |} Is it ever important for a user to be able to understand their model? Why should one care to interpret it?
When we do care, it is for three primary reasons: 
\begin{itemize}
    \item \textbf{Trust}: Interpretability is a necessary aspect of trusting and validating a model. This is such an important point that will discuss it in much more detail below.
    \item \textbf{Downstream explainability}: Interpretability can greatly aid in, and is sometimes necessary for, achieving an explainable model. A model that one can understand is one that outputs useful and generalizable predictions and depends on parameters grounded in domain knowledge. We will explore this further in  \Sec{explainability}.
    \item \textbf{Computational efficiency}: Interpretability often comes with gains in computational efficiency. An interpretable model is, by definition, easily approximated by a human, and this is usually highly correlated with being easily approximated by a computer. 
    Simple functional forms can often be run quickly and efficiently, and it is common for uninterpretable ML taggers to be ``distilled'' into fast and interpretable ones for use in high-demand environments like the LHC trigger system~\cite{Duarte:2018ite, Bal:2023bvt, Schulte:2025mai}.
\end{itemize}

\paragraph{Trust and uncertainties |}

\textit{Trust} relates to an understanding of the model uncertainties of all types.
There are many types of uncertainties an ML model can have, some of which are easy to quantify and some of which are difficult to quantify.
All of these types of uncertainty are reduced with increased interpretability.
We present examples of these types of uncertainty in the form of questions one should ask themselves of their model:

\begin{itemize}
    \item \textbf{Robustness}: The data available for training is inherently finite and noisy, and the procedures used to estimate model parameters are necessarily approximate. To what extent is it acceptable for parameters to deviate from their optimal values or exhibit bias, and which parameters are most susceptible? Is the model robust to adversarial perturbations, and are all parameters equally informative? These questions are more readily addressed with an interpretable model. Using Newton's Law as an example, even if the value of $G_N$ were incorrectly determined, one would still predict stable orbits as a robust model property. 
    \item \textbf{Inductive bias}: Neural networks $f_\theta$ can only ever \emph{approximate} the ``true'' model $f$, and they will frequently be biased. While it is often the case that model inductive biases, such as the famous spectral bias of neural networks that promotes smoothness~\cite{rahaman2019spectralbiasneuralnetworks}, are often good modeling choices for most datasets, this is not always guaranteed. To what extent do these approximations deviate from the true model, and how close are the two in practice? Answering this question requires an interpretable model and interpretable performance metrics. 
    
    \item \textbf{Domain uncertainty}: What is the domain of validity, i.e.\ the set of inputs for which the uncertainty in $f_\theta(x)$ remains acceptably small? Are there specific inputs that yield undesirable or unreliable predictions? For example, by understanding the form of Newton's Law, one would know to avoid the regime where $r \to 0$, whereas this would remain unclear in a less interpretable model. 
    \item \textbf{Verifiability / transparency-informed priors}: 
    How can one effectively communicate the structure and behavior of a model to others? If presented with a model from another source with minimal documentation, how can its function and the meaning of its outputs be reliably understood? Interpretable models facilitate transparency, reproducibility, and trust, making it easier to share and validate results across different users and contexts.

    \item \textbf{Intervention}: 
    Once the model has been interpreted, what actions can one take based on that understanding? What is the purpose of obtaining a simpler approximation? Are there expectations informed by domain knowledge, and how should deviations from these expectations be addressed? If one computes interpretability metrics, such as Shapley values, will these metrics meaningfully inform downstream decisions or interventions? If not, reconsider whether interpretable structures or metrics are necessary at all.
    Note that this highly intersects with ``explainability for validation'', and requires establishing a decision or intervention procedure ahead of time to avoid confirmation bias.
    We discuss this further in \Sec{task}.

\end{itemize}

\section{Explainability}
\label{sec:explainability}

Explainability concerns our ability to \emph{extract knowledge} from a model.
This is done by associating the model (or sub-components of the model) to domain knowledge.
Oftentimes, this domain knowledge takes the form of a simpler and more interpretable physical model, a sub-component of a model, or a fundamental law, principle, or relation. 
Importantly, \emph{explainability does not exist without domain knowledge to compare to}. 
In other words, unlike interpretability, there is no definition of ``explainability'' in isolation.
That explainability in models requires previous models or domain knowledge to compare against connects to broader ideas in neuroscience, philosophy, and mathematics that \emph{knowledge is inherently relational}, that is, scientific or empirical information is meaningless except in the context of other information.

In this section (in a similar fashion to the previous section about interpretability), we aim to explore the spectrum of explainability, its tradeoffs with data adaptability, and pose questions about the necessity of explainability.
Since interpretability and explainability are intertwined, and it is not possible to achieve explainability without interpretability, much of the discussion here will be correlated with the discussion in the previous section.
The key distinguishing feature to keep in mind therefore, is that \textit{explainability is about the scientific content of the model as it  relates to domain knowledge and data}, whereas interpretability is solely about the science-agnostic structure of the model.

\paragraph{Explainability is not just for ML |}
Explainability is a fundamental consideration not only for ML models; it is in fact a core challenge for all modeling, and even human-derived classical models can fail to achieve full explainability. 
For example, cosmology and astrophysics often rely on models which are entirely based on empirical functions (e.g.\ a double-power law) with free parameters often calibrated to simulations. 
These models include the density profiles of dark matter halos \cite{NFW, Einasto1965}, or the halo mass function \cite{Tinker:2008ff, Despali:2015yla}, both of which are central ingredients in cosmological analyses. 
These models are interpretable: we understand how their parameters shape the output and know its functional form. 
However, we cannot explain the origin of their specific shape, the specific parameter values or emergent patterns (such as universality) found in simulations. 
In particle physics, one also encounters incomplete explanations.
Hadronization models like those found in \textsc{Pythia}~\cite{Bierlich:2022pfr} involve parameters that cannot be derived from first principles and are largely phenomenological.
Even the origins of the parameters of the SM are unexplained, and frameworks such as Standard Model Effective Field Theory (SMEFT) or the $\kappa$-framework~\cite{10.1093/ptep/ptac097} deliberately sacrifice first-principles explanations in favor of data adaptability.
\subsection{Spectrum of Explainable Models and Tools}\label{sec:spectrum_exp}
\begin{figure}
    \centering
    \includegraphics[width=\linewidth]{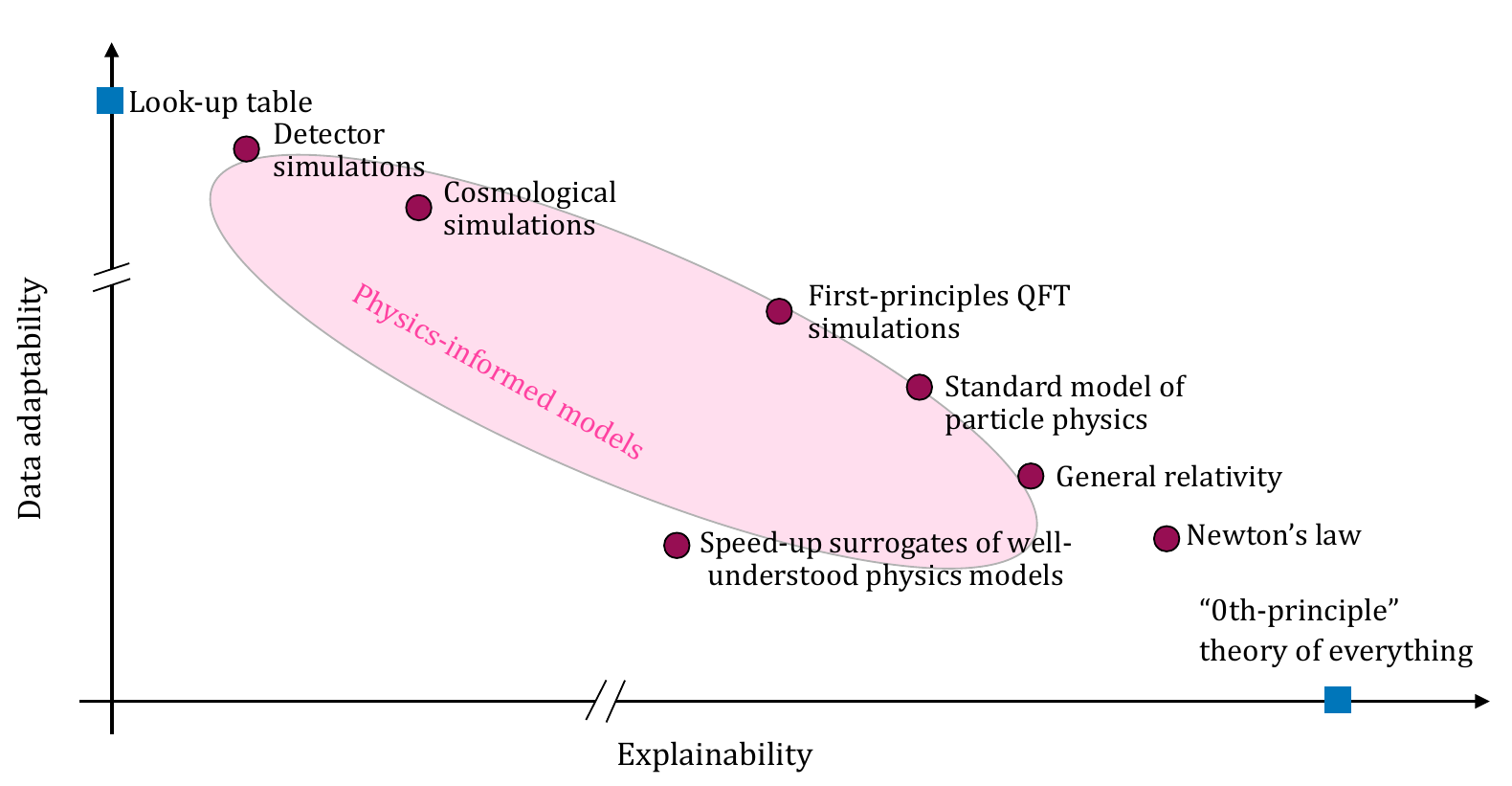}
    \caption{The trade-off between explainability and adaptability:  the ability to model the full data distribution. Models designed for explainability tend to focus on broad, high-level patterns in the data, often at the expense of accurately representing the complete complexity of the data. We represent existing physical models with circles and extreme examples with squares. We represent ``physics informed models'' with a large pink blob, representing the wide breadth of explainability in physics-informed models.}
    \label{fig:graphics_exp}
\end{figure}
Explainability, and the extent to which models yield extractable scientific information, is, unsurprisingly, a wide spectrum. 
We explore this spectrum in \Fig{graphics_exp}.
Here, the horizontal axis ranks expressivity broadly, and the vertical axis ranks the ability of the model to fully fit an arbitrary data distribution, which we will detail more precisely below.

\paragraph{Maximally unexplainable: ``not even wrong'' |}

As with interpretability, it is pedagogically useful to construct a maximally unexplainable model:
For supervised learning tasks with data pairs $(x,y)$, a model that perfectly minimizes most losses is just a simple look-up table of the data.
This matches the \emph{fitting} data perfectly, yet it contains zero predictive power for \emph{unseen} data. 
In other words, no information about the underlying full data distribution can be extracted.
The fact that it cannot make any generalized prediction about the underlying data distribution is what makes the model maximally unexplainable. 

\paragraph{Maximally explainable: ``theory of everything'' |}
On the opposite side of the spectrum, one can imagine a maximally explainable model: a hypothetical, parameter-free ``zeroth-principle model'' (in contrast to a first-principles model) or ``theory of everything''\footnote{Not in the string landscape sense, which has too much freedom.} which includes the full Lagrangian of the universe.
Such a model would be maximally explainable since all other prior domain knowledge could be derived from it, in the sense of a top-down fundamental theory.

\paragraph{In between: a broad range | }In the middle are more realistic scenarios including both ML-based models, leaning to the unexplainable part of the spectrum, and classical models, leaning towards the explainable side.
It is much more difficult to broadly classify the relative explainability of these models, as explainability cannot exist outside the context of domain knowledge for a specific scientific problem.
We exemplify the range of explainability in models with varying degrees of physics information injected as the broad pink blob in \Fig{graphics_exp}.
In many cases, however, the barrier to explainability is interpretability, allowing for an approximate ranking of explainability in models using interpretability as the major determining factor.
In order to associate a feature of a model with domain knowledge, one must be able to first interpret that feature. 
An interesting example is what we call ``second-principles'' models, which are phenomenological models that are partially explainable in terms of physics principles and are usually simple to interpret, but have not been fully connected to a first-principles physics model.
Examples include prescriptions for star formation, supernova feedback, and AGN feedback used in hydrodynamical simulations \cite{2020NatRP...2...42V} and phenomenological hadronization models~\cite{Andersson:1986au} used in simulating collider events. 

\subsection{Trade-offs: Explainability vs. Adaptability}

\paragraph{Data adaptability |}

Like with interpretability, there is a fundamental ``cost'' to explainable models from which scientific knowledge can be extracted: this cost is related to the model's ability to fit arbitrary datasets, or to ``adapt'' to qualitatively different data, shown as the vertical axis in \Fig{graphics_exp}.
A model's adaptability to data is defined by the breadth of data distributions it can fit with its free parameters.
Adaptability is related to and intertwined with, but subtly different from, expressivity, in the same way that explainability is highly intertwined with interpretability.
Adaptability (associated with explainability and scientific content) is a relationship between a model and data, concerning whether the model is compatible with the data, while expressivity (associated with interpretability and model structure) is an inherent property of the model design alone.
Just as more interpretable models are generically more explainable, more expressive models are generically more able to fit arbitrary data.

Adaptability is useful in the context of model comparison.
One can compare a more explainable model to a more adaptable model on a test set to determine if the data can be adequately explained by the former model, or if a more complex or different explanation is required.
For example, the SM can only describe Poincar\'e invariant data, but it cannot fully fit cosmological data as this data is not  globally  Poincar\'e invariant, necessitating a more data-adaptable cosmological model.

\paragraph{The trade-off |}
It is instructive to first look back at the extreme cases presented in \Sec{spectrum_exp}.
A model that can represent arbitrary input-output mappings trivially, such as the look-up table as shown in the top-left corner of \Fig{graphics_exp}, admits only point-wise explanations -- its output is simply the stored value for a given input -- but does not admit meaningful global explanations of its behavior. 
On the other hand, a look-up table can be used to fit \textit{any} arbitrary dataset simply via pure memorization.
In a sense, a look-up table is ``not even wrong'', as it is infinitely adaptable and can realize arbitrary datasets, but with no compressible structure to enable explanations. 
That is, one cannot learn anything or make predictions from a look-up table.
On the other extreme, the fully specified ``theory of everything'' with no free parameters has zero ability to adapt, as its fully fixed functional form only applies to the dataset of interest, and will be incorrect for any other arbitrary dataset. 
This is a case of minimal adaptability and maximal explainability as it allows for a complete global description of its behavior, but with no flexibility to explain any other hypothetical data.

The ability to fit arbitrary data is not always desirable, as explainability requires some structure.
Domain knowledge or explicit inductive bias is needed to restrict the space of solutions beyond the whims of training dynamics: on the higher end of the explainability spectrum are models that incorporate domain knowledge such as symmetry, semantically meaningful parameters, and priors.
This applies not only to ML models, but also to physical theories.
For example, the SM incorporates broad physical principles (Poincar\'e symmetry, locality, and unitarity), semantically meaningful parameters (masses and couplings), and even physical priors (to taste, one's opinions about Higgs naturalness and strong-CP beyond the SM). 
The cost of this explainability is the need for assumptions and the risk of mismodeling: if the universe were not (locally) Lorentz invariant, the SM could not model it.
The same is true in cosmology. A central pillar of the standard model of cosmology ($\Lambda$CDM) is the cosmological principle, i.e., the assumption that on sufficiently large scales the universe is homogeneous and isotropic. If this were not true, $\Lambda$CDM would break down.
When training large models, inductive bias will always be a concern.
It is usually best to guide the inductive bias through domain knowledge and model design (e.g.\ the physics-informed structure of \Sec{intrinsic}) during task specification (\Sec{task}) than to let numerics, gradient descent, random initialization, and spectral bias decide the network output for you.

\subsection{When Should One Care About Explainability?}\label{sec:care_explainability}

\paragraph{When not to care: never! |} Are there any applications that do not require explainability? In contrast to interpretability, we argue that some degree of explainability is almost always necessary in scientific contexts, and that it should not be de-prioritized.
Consider, for example, a setting in which ML is used as a fast surrogate model (e.g.\ a tagger trained to mimic a simulation). Here, the primary goal is to achieve the highest possible accuracy with minimal computational cost. In such cases, having a simple, interpretable functional form is less critical; the model simply needs to perform well. Explainability, however, remains essential: careful validation and interrogation of the model are crucial in order to understand and articulate when, how, and why the model performs reliably in its downstream application.
Phenomenological or ``second-principles'' models, like the Lund string model for hadronization~\cite{Andersson:1986au} or effective field theories, often sacrifice some degree of explainability in order to adapt to data, but even these models are still explainable and, importantly, can at least match onto existing models or in principle become explained by new models.

\paragraph{When to care: validation, prediction, and discovery |} When one does care about explainability, it is usually for three reasons: \emph{validation}, \emph{prediction}, and \emph{discovery}. 
As described in detail below, all three of these principles are essential for scientific models, which is why explainability is almost always indispensable.  

\paragraph{Validation, trust, and uncertainties |} The purpose of ``explainability for validation'' is, similar to interpretability, to improve trust in the model and reduce uncertainty. This can be achieved by manipulating the data -- for example, testing various noise levels -- or by interrogating the model under different limits and regimes.
If one uses Shapley values or related metrics (explored more in \Sec{posthoc}) to rank the relative importance of input variables, and this ranking matches expectations derived from domain knowledge, then one's trust in the model can increase, since the model is similar to an already-trusted model.
This can, of course, potentially be sensitive to the choice of metric -- for example, Shapley values have a precise game-theoretic definition (\Eq{shapley}) and it is rare to compute these metrics in the original classical theory quantitatively, but nevertheless these qualitative proxies can reduce qualitative
uncertainties.
Similarly, if a model has a constrained latent space (due to symmetry, a special loss, disentanglement, or other constraints), these parameters can also be approximately mapped onto the original model.
This too, can increase model trust.

\paragraph{Prediction |} Often, the goal of a scientific model is to either interpolate or extrapolate to predict unseen data, leading to ``explainability for prediction''. 
Given a training dataset that we fit a model to, we would like the model to \emph{generalize} and make accurate predictions outside the training set -- for instance, predicting the time and location of a potential future eclipse using past astronomical data as a training set.
Prediction inherently requires a constrained model (that is, a model of limited adaptability): if a model could have predicted anything, no information is gained.
An unexplainable model, like a look-up table or an overfit model, cannot generalize or make meaningful predictions outside its training set, as they are unconstrained.
An explainable model, in contrast, is constrained by either physical principles or known inductive biases.
If the inductive biases are unknown (e.g.\ the model is uninterpretable), it is not possible to tell if the model will truly generalize or not.\footnote{A black box model with inductive biases enforcing smoothness \emph{might} generalize on a specific dataset, but it is impossible to tell if this is a statistical fluke, if this holds true for the full data phase space, if it will still hold if the data is slightly modified, etc.}

It is often claimed that highly overparameterized (and therefore, naively highly unexplainable) ML models can still generalize well~\cite{DBLP:journals/corr/ZhangBHRV16}, in an apparent contradiction of classical statistical intuition, such as in double descent~\cite{Belkin_2019, nakkiran2019deepdoubledescentbigger} or benign overfitting~\cite{Bartlett_2020}. 
We emphasize, however, that after a full accounting of all inductive biases and data-dependent statements, the apparent contradiction disappears. 
First, the training or fitting procedure (which is itself part of the model definition) often has an implicit regularization, which typically selects for smooth or minimum-norm solutions as an inductive bias and constrains the model~\cite{gunasekar2017implicitregularizationmatrixfactorization, soudry2024implicitbiasgradientdescent}. 
These regularizations are modeling choices (often unintentional), but can in principle be chosen with intentionality.
Second, generalization is an inherently dataset-dependent statement (in the same way that explainability is).
It does not make sense to say a model generalizes -- it only makes sense to say a model generalizes on a specific dataset or group of datasets, and this is a statement that the model's inductive biases (either unintentionally, or by design) is a good descriptor for the data.
It is often the case that many real-world datasets (cosmological data, images of dogs and cats, etc.) and toy datasets comprised of simple mathematical operations are well-described by smooth low-frequency interpolation or simple models~\cite{Semenova_2022}, explaining the success of neural networks in generalization on these datasets~\cite{rahaman2019spectralbiasneuralnetworks, John_Xu_2020}, \emph{but this is not guaranteed}, as there exist pathological datasets or tasks that are badly misaligned with typical inductive biases.
Thus, \emph{trusting} predictions and generalizations using ML, especially in the overparameterized regime, often requires explanation to relate inductive bias to domain knowledge about the dataset. 

\paragraph{Discovery |} We may hope to gain new scientific information from our models, which leads to ``explainability for discovery''.
In cases where input features are assigned physical meaning, tools related to feature importances (such as Shapley values or others) can do more than diagnose model behavior -- they can reveal how different physical inputs determine the final outputs of interest. 
This may confirm expectations from domain knowledge, or reveal new patterns that were previously unseen with more standard statistical techniques (e.g.\ \cite{Lucie-Smith:2019hdl, Lucie-Smith:2022mar}). 
For example, suppose a symbolic regressor extracts $V(r) = \frac{-GMm}{r} + \frac{\alpha}{r^3}$ from gravitational data. 
From the perspective of a pre-1920s physicist such as Isaac Newton, one must decide whether to (a) attribute this to a deficiency in the simplicity metric, (b) reduce trust in the model due to its unexpected form, or (c) regard it as evidence that Newtonian gravity is incomplete. 
This is no different from the reasoning that is applied to any new scientific discovery, regardless of whether it was driven by a classical or ML model.

Disentangled latent representations may also yield new discoveries.
The disentangled latent representation learned by a network can be thought of as the non-linear, underlying degrees of freedom for the problem at hand. 
Their structure and information can be quantified and interpreted within the specific scientific context, by connecting them to other physical quantities of interest via information-theoretic metrics. 
This in turn allows one to identify which latent variables encode physically relevant quantities. 
Such analyses have been done in several works to identify the independent degrees of freedom in latent spaces for observables like the halo mass function \cite{Guo:2024had} and density profiles \cite{Lucie-Smith:2024xsx, Lucie-Smith:2022uvv}, and to reveal new connections between the dark matter halo structure and evolution~\cite{Lucie-Smith:2023kue} and the impact of early dark energy on the cosmic microwave background~\cite{Piras:2025eip}.
The resulting ``knowledge extraction'' approach opens the door to the potential of machine-driven new scientific discoveries.

\section{Methods: Intrinsic and Post-Hoc}\label{sec:intrinsic_interpretability}

There are two principal strategies for achieving interpretability and explainability in models: \emph{intrinsic} methods, in which desired structures are embedded into the model by design, and \emph{post-hoc} methods, in which diagnostics are extracted from an already-trained model's structure. 
Most of these methods are directly tied to interpretability and only indirectly tied to explainability -- often, they only offer explainability by first achieving interpretability.
Thus, we will frame our discussion in terms of interpretability, keeping in mind that often the goal of interpretability is downstream explainability.
Intrinsic and post-hoc methods are distinguished by \emph{when} interpretability is achieved -- either before or after the model is fixed.
Below, we describe representative techniques from each category.

\subsection{Intrinsic (By-design) Approaches}
\label{sec:intrinsic}

The central conceit of interpretability-by-design is that one has a choice (and a responsibility) in selecting and designing models, and therefore can directly select and design based on what one considers interpretable. 
Classical models, often in the form of parametric fits, are constructed with this principle, but this can easily be extended to ML.
There are few ``off-the-shelf'' methods for intrinsic interpretability: physics-informed structures and representation learning are problem-dependent by definition, and tools like principal component analysis, classical ML, and symbolic regression are quite broad in their applications.

\paragraph{Physics-informed structures |} Physics-informed models introduce a degree of interpretability by embedding known physical structures into the learning process. 
This often takes the form of imposing structure on ML architectures or loss functions, and is usually achieved by either generalizing parts of an existing physical model (e.g.\ generalized event shapes~\cite{Ba:2023hix}, generalized Lund fragmentation~\cite{Bierlich:2024xzg}, and modular, rotationally equivariant set-based architectures~\cite{Thiele:2022}), building a universal function approximator that respects specific physical properties (e.g.\ infrared and collinear safe networks~\cite{Komiske:2018cqr}, Lorentz-invariant networks~\cite{Spinner:2024hjm}, manifest perturbative structure~\cite{Gambhir:2025lka}, and rotation/reflection-equivariant galaxy classifiers~\cite{Dieleman:2015,Pandya:2023,Scaife2021}), or deriving a loss whose minima or training dynamics encodes those same properties (e.g.\ enforcing symmetries~\cite{Hebbar:2025adf}, solving PDEs~\cite{raissi2017physicsinformeddeeplearning,mishra2025spinnadvancingcosmologicalsimulations, Dai:2024}).
A broader approach is to place regularization or priors on model parameters, as models often simplify in limits: for example, an L1 regularizer can force parameters to be sparse and an L2 regularizer can force parameters to be close to zero, which can simplify a model. 
Often, these methods are (relatively) highly explainable compared to ordinary ML models, as direct domain knowledge has been imbued into the structure and therefore physical properties (such as symmetries, specific important parameters, or useful encodings) can either be guaranteed or easily read off.
However, since they typically rely on flexible models such as neural networks, they are not as intrinsically interpretable as simple parametric models, and may still benefit from additional interpretability or explainability tools.

\paragraph{Representation learning |}
Representation learning involves mapping high-dimensional data into a lower-dimensional ``latent'' space. 
A classical example of this is \emph{principal component analysis} (PCA), which learns a linear dimensionality reduction of the input data. The principal components are orthogonal and ranked by variance, making PCA inherently interpretable.
Deep learning generalizes this idea to nonlinear mappings, potentially revealing hidden generalized structure. 
Representation learning offers the opportunity to design new models that can turn black-box architectures into interpretable models, if the generated representations can be made interpretable and/or explainable. 
Disentangled representations, in particular, aim to isolate distinct, independent factors of variation in the underlying data within different, independent latent dimensions. 
For instance, methods like InfoGAN \cite{Chen2016InfoGAN} maximize the mutual information between a subset of latent variables and certain labels, encouraging each such variable to capture a distinct and semantically meaningful factor of variation.
The $\beta$-VAE \cite{Higgins2017betaVAE} takes a complementary approach, promoting disentanglement by balancing reconstruction accuracy with latent space independence. 
Without explicit inductive bias on the model architecture, data, or training objective, it has been argued that informative disentangled representations are difficult to achieve~\cite{locatello2019challengingcommonassumptionsunsupervised}. 
In practical examples that include inevitable inductive biases, others have demonstrated that encouraging orthogonality between latent dimensions leads to more interpretable representations. Building on these ideas, the SciNet model~\cite{Iten:2020ohp} showed that the learned latent parameters recover the expected physical parameters of the problem, with further success in cosmological applications of the interpretable variational encoder (IVE) model~\cite{Lucie-Smith:2024xsx, Lucie-Smith:2022uvv, Piras:2025eip, Guo:2024had}.

\paragraph{Traditional ML |} Simpler ML models, such as linear regression or ensembles of decision trees, tend to be easier to interpret.
One can directly inspect which input features drive outputs via linear coefficients or binary tree splits. 
While deep learning has superseded them for many high-dimensional tasks requiring greater expressivity, these methods continue to serve as valuable interpretable baselines and are often the right tool when the priority is transparency over raw performance.

\paragraph{Symbolic regression |} Symbolic regression is the procedure of approximating either data or a larger model with a symbolic expression. 
This includes ordinary parametric fitting, where a functional form (or family of functional forms) is specified ahead of time.
These methods typically have a set of pre-defined mathematical operations allowed between symbols, and a notion of the complexity of an expression.\footnote{An ordinary parametric fit can be viewed as a special case of symbolic regression with a very stringent complexity budget.} 
A popular tool for symbolic regression is PySR~\cite{Cranmer2023}. 
In practice, the interpretability-expressivity trade-off discussed in \Sec{interpretability_trade_off} can be controlled through the choice of complexity penalty.

\paragraph{Dictionary learning and sparse encoding |}
Similar to PCA, which attempts to find an ordered linear basis to approximate the data, dictionary learning~\cite{OlshausenField1996, Aharon2006KSVD, Mairal2009} attempts to find an \emph{overcomplete} basis (``atoms'') that form sparse linear combinations to reproduce the data. 
This can also be performed on a model's internal data representation, called a \emph{sparse autoencoder}~\cite{Ranzato2006, Ng2011sparse, Makhzani2014ksparse}, to see if there are simple linear compositional features.
Each data point is a linear combination of a handful of atoms from a large repertoire.
This is especially useful as a post-hoc technique for LLMs~\cite{cunningham2023sparse, bricken2023monosemanticity, templeton2024scaling, gao2024scaling}: sparse encoding is a good inductive bias for language, as sentences can be easily decomposed into combinations of concepts from a large dictionary.

\paragraph{Chain-of-thought and reasoning traces |} For LLMs, chain-of-thought (CoT)\cite{wei2022chain} prompting gives the model an ``internal monologue'' that allows the model to break problems into steps and check its own work in a way that somewhat mimics human reasoning.
The CoT can often be directly viewed, providing an interpretable window into the LLM's reasoning process, which can be useful for \emph{intervening} (more in \Sec{intervention}) if an error occurs, or can be used to post-hoc-explain the LLM's output given its input and training data.\footnote{Here, we are careful to frame LLMs as ``models'' insofar as they predict tokens given a parameterized fit to data like all other models considered here, rather than framing them as ``agents'' performing actual calculation or reasoning. In this view, embedding an LLM with CoT is an intrinsic modeling decision that adds additional structure.} 
Care should be taken when interpreting CoT traces, as they do not necessarily faithfully reflect the underlying model computations\cite{turpin2023language, lanham2023faithfulness}.

\subsection{Post-Hoc Techniques}
\label{sec:posthoc}
Post-hoc interpretability methods treat the model as a black box to be opened or probed.
Rather than directly adjusting the model design as with intrinsic methods, post-hoc methods are typically model-agnostic and probe the structure of an already trained model.
Often, these are ``off-the-shelf'' methods and are not tuned for a specific problem or scientific task, and their outputs should be considered with this in mind.

The two major classes of post-hoc techniques are \textit{metric-based} and \textit{distillation-based}.
Metric-based tools involve assigning a number, usually with some well-defined statistical interpretation, to aspects of a model.
These metrics usually characterize the relative importance of different aspects of the inputs $x$ or the parameters $\theta$ to the model output $f_\theta(x)$, or some notion of the performance of the model on a particular task.
Distillation-based methods instead aim to approximate the black-box model with a simpler, more interpretable surrogate.

\paragraph{Shapley values | }
The most widely-used metric-based interpretability tool is the Shapley value~\cite{Shapley:1953}, borrowed from cooperative game theory.
Shapley values are used to assign ``importance'' to features, serving as a tool to assess which input variables significantly contribute to a model's predictions.
The Shapley value $\phi_i$ of feature $i$ in the model $f$ is defined as the average marginal contribution of feature $i$ across all possible orderings in which features can be added:
\begin{align}
\phi_i(f) = \sum_{S \subseteq \mathcal{F} \setminus \{i\}} \frac{|S|!\,(|\mathcal{F}|-|S|-1)!}{|\mathcal{F}|!} \left[ f_{S \cup \{i\}}(x) - f_S(x) \right],
\label{eq:shapley}
\end{align}
where $\mathcal{F}$ is the full set of features, $S$ is any subset not containing feature $i$, and $f_S$ denotes the model evaluated with only the features in $S$ (with the remaining features marginalized over).
SHAP~\cite{lundberg2017unified}\footnote{\href{https://shap.readthedocs.io}{https://shap.readthedocs.io}} is a widely-used numerical implementation of Shapley values for ML models.
It is common to use SHAP to evaluate taggers~\cite{Bhattacherjee:2022gjq, Vent:2025ddm, Patel:2026zbq}, though as will be discussed in \Sec{intervention}, care should be taken to ensure these metrics are meaningful in the context of a full analysis plan.

\paragraph{Saliency maps |}
Saliency maps provide a pixel-level attribution for image-based models by computing the gradient of the model output with respect to each input pixel~\cite{simonyan2014deepinsideconvolutionalnetworks}: large gradient magnitude indicates that the model's prediction is sensitive to that region.
Since vanilla gradients tend to produce noisy maps, several extensions, including Class Activation Mapping~\cite{zhou2015learningdeepfeaturesdiscriminative}, Grad-CAM~\cite{Selvaraju_2019}, and SmoothGrad~\cite{smilkov2017smoothgradremovingnoiseadding}, may be used to improve spatial smoothness and class specificity by averaging over perturbed inputs or pooling over intermediate feature maps.
Saliency maps characterize model sensitivity (a model's structure) rather than causal importance (which would be the realm of explainability, and require additional modeling), and only capture linearized relationships.
\paragraph{LIME |} Local Interpretable Model-Agnostic Explanations (LIME)~\cite{ribeiro2016should} generates an interpretable local approximation to a black-box model in the neighborhood of a single prediction, akin to a Taylor expansion.
Given an input $x$ of interest, LIME perturbs $x$, collects the model's responses on the perturbed samples, and fits a simple surrogate model (typically linear regression) weighted by the proximity of each perturbed sample to $x$.
The surrogate model is not a global description of $f$ (it is only valid locally) but it provides an interpretable surrogate near $x$.
The choice of perturbation scheme and proximity kernel are user-defined hyperparameters that can significantly affect the resulting explanation, and should be chosen with the structure of the problem in mind.

\paragraph{Distillation |} Many post-hoc methods involve approximating a larger model with a simpler model. 
Knowledge distillation~\cite{hinton2015distillingknowledgeneuralnetwork}, or the ``student-teacher'' approach, involves training a smaller ``student'' ML model to mimic the output of an existing ``teacher'' model.
The student model is itself often chosen to have an intrinsically interpretable-by-design structure, employing the methods described above in \Sec{intrinsic} -- that is, the student's structure \emph{is} the interpretation.
A common example is \emph{symbolic distillation}, where the student is an analytic expression found via symbolic regression~\cite{Cranmer2023}.
Distillation is also often used in environments where model speed and size are limiting factors, such as the LHC~\cite{Liu:2023dio, Bal:2023bvt}.

\paragraph{Probing |} A \emph{probing classifier} is a simple model trained to predict a property of interest from a network's internal representations~\cite{alain2016probing, belinkov2022probing}, to determine what information is potentially stored in the representation.
For example, to check if a model trained on collider physics data is potentially using the mass of a jet in its answer, a second model can be trained using an internal layer of the original model as input to see if it can regress to the jet mass.\footnote{In fact, this trick is used in \emph{mass-decorrelated taggers}~\cite{Shimmin:2017mfk}, where the goal is to ensure that the network does \emph{not} know about jet mass.}
Note that if a probe finds the feature is represented, that does not necessarily imply the network uses the feature downstream or that it is \emph{efficiently} represented in the latent structure\cite{hewitt2019control}, and conversely, if a probe fails to find a feature, that may only indicate a failure of the probe's training.\footnote{See \Reference{Rai:2026uzm} for an example application to jet tagging.}

\section{Considerations: Tasks and Interventions}
\label{sec:task}

We identify two key actions to consider when accounting for the interpretability and explainability of a model: (1) \emph{task specification} and (2) \emph{intervention plans}.

Proper task specification can aid in avoiding modeling pitfalls: while it is common in jet physics to aim for as high a classification performance as possible, tagging is only an intermediate step in the true downstream task of parameter inference,\footnote{In fact, jet labels are often not physically meaningful in the first place. See \Reference{Komiske:2018vkc}.} and much of the apparent extra performance in classification washes out at the final inference step~\cite{Gambhir:2025xim}.
The intervention plan is therefore part of the model design.
In this light, the decision to use post-hoc methods, when accounted for in the analyses plan, is indeed just another aspect of intrinsic model design.

\subsection{Task Specification} Task specification is the precise determination of what one's model must accomplish and what properties it must satisfy. 
In practice, the desired properties of a model -- e.g.\ predictions with controlled uncertainties or practically useful information compression -- are not directly codified by simple loss function choices and model section, and so care should be taken to ensure the model accomplishes what we really want.
Concretely, this means specifying:
\begin{itemize}
    \item \textbf{The downstream scientific application}: What questions are the model ultimately being used to answer?
    Often, we deal with surrogate models for surrogate problems. 
    For instance, are we training a classifier because we need to assign labels to observed data for its own sake, or are these labels on data itself just a step in some final measurement or parameter estimation? What is physical and what is a proxy?

     \item \textbf{The desired properties:} Does one need controlled uncertainties, compression, robustness to distribution shift, or physically meaningful parameters? 
    What does one want the model to learn, and what does one \emph{not} want it to learn?
    These desires are often not directly codified by a simple loss function.
    If all one wanted to do was minimize a mean-squared error, a simple look-up table would have sufficed.
     \item \textbf{The degree of interpretability and explainability:} Based on the discussions in \Secs{interpretability}{explainability}, what level of structural transparency and domain-knowledge alignment is required for the task at hand? 
    For instance, if one were performing a precision $\alpha_s$ extraction, explainability is crucial, while interpretability is not.
    \item \textbf{Implicit modeling choices:} Architecture selection introduces inductive bias -- for example, the choice to use an MLP over a transformer induces implicit regularization and spectral bias that should be checked against domain knowledge. 
    These choices are part of the task specification even when they are not explicitly recognized.
\end{itemize}
Task specification is essential for integrating intrinsic interpretability or explainability, since one must know what properties are desired \emph{before} embedding them into the model design or choosing a post-hoc strategy.

\subsection{Intervention Planning}
\label{sec:intervention}

Intervention plans concern what will be done with an interpreted or explained result.
If no downstream decision, such as a change in uncertainty or trust, a rejection or acceptance of a model, or a modification to the analysis pipeline, follows from having interpreted or explained a model, one should question whether interpretability or explainability was needed in the first place. Concretely, an intervention plan means specifying:

\begin{itemize}
    \item \textbf{Downstream decisions:} What would cause one to either gain faith in or lose trust in the model? Under what circumstances is a surprising result a model failure, a bug, or a discovery? Under what circumstances is an \emph{unsurprising} result a model failure, a bug, or a confirmation? What will warrant changing or discarding the model? 
    \item \textbf{Model transparency:} What does one need to understand about their model for making downstream decisions? Are these inherent structures or post-hoc metrics and distillations? For example, is there a particular SHAP value that a jet tagger might return for a particular feature that would cause one to change their approach to the tagger? Would a relative feature suffice, or is one comparing to a theoretical SHAP value computed in some base model? 
    \item \textbf{Actionable items:} What can be done with the model after having been interpreted or explained? If, for instance, a SHAP value behaves as expected, will the uncertainty on the model quantitatively decrease, or merely qualitatively? Can the model be discarded or changed, and should it? Is the model easy to debug?
\end{itemize}

\paragraph{Intervention and avoiding confirmation bias |}Intervention planning is especially important with regard to confirmation bias.
Without a pre-specified plan, interpreted results are susceptible to post-hoc rationalizations in both directions.
If a SHAP value, for instance, shows that a model relies on expected input features, this only confirms that the model is consistent with prior expectations, not that it is correct.
Conversely, if unexpected dependencies are revealed, one must decide to reject the model, investigate further, or regard the result as a discovery.
This decision should not be made after seeing the result, in much the same way experimental analyses are blinded. 
One should avoid the temptation to apply post-hoc methods without an intervention plan, as otherwise there is little value to be gained from them.
We emphasize that none of this discussion is specific to ML models, and task specification and intervention planning applies equally to all scientific tasks and modeling.

\section{Conclusion}
\label{sec:conclusion}

In this review, we have aimed to elucidate the nature of interpretability and explainability for ML applications in physics.
The ability to understand the model mechanically (\emph{interpretability}) and the ability to relate the model to scientific knowledge (\emph{explainability}), whether achieved intrinsically or post-hoc, are choices to be considered when building an analysis pipeline for a scientific task.
We emphasize that interpretability and explainability are not always necessary, and come with trade-offs in terms of model expressivity and adaptability to generic data, and thus it is recommended to always consider why one may need interpretability/explainability in their models and what precisely one will do with it if they have it.
If nothing else, one should take away the importance of \emph{task specification} and \emph{intervention planning} as explicit steps to consider in the model design process.
While there are a number of tools on the market for post-hoc interpretability and explainability, we have also aimed to highlight intrinsic methods through explicit model design and parameterization as a powerful method for control of models.

Throughout, we have emphasized that ML models are ``nothing new'', in that classical physical models are also subject to the exact same questions of interpretability and explainability.
The only difference is scale, and the features we generally associate with classical models, such as semantic meaning for parameters, are simply consequences of the increased interpretability or explainability of classical models.
Any scientific question one may ask of classical models is equally valid to ask of a ML model, and vice-versa.

Interpretable and explainable modeling will continue to gain interest in the physics community as the number of tools and ease of use increase.
Along these lines, it is worth asking whether interpretable and/or explainable AI has been useful thus far.
The answer is clearly yes: not only have we learned a significant amount of physics with ML, but this physics \emph{could not have been discovered without ML}. 
Just jet-tagging at the LHC alone has accounted for significant improvements in measurements and discovery potential at colliders~\cite{Mondal:2024nsa, hepmllivingreview}, which requires taggers to be interpretable and explainable in order to be validated, calibrated, and matched to SM expectations. 
Studies of cosmological structure formation, where explainability was set as a priority, have revealed new physical insights otherwise difficult to extract with traditional methods -- for example, on the origin of late-time halo properties \cite{PhysRevD.109.063524, Lucie-Smith:2023kue}, a previously unknown self-calibration mode for cluster surveys \cite{Ntampaka_2022}, and the identification of which features in weak-lensing maps carry non-Gaussian cosmological information \cite{Matilla2020, PhysRevX.12.031029}.
These results demonstrate that treating interpretability and explainability as design requirements  can turn machine learning into a genuine instrument of scientific knowledge.

Another interesting question is whether we have learned anything \emph{from} (not with) ML -- that is, has explainable AI actually explained anything novel?
In HEP, the answer seems to be ``not yet''. 
While bounds placed using ML analyses techniques help us restrict physics, and model building can be automated with the help of AI tools, much of the progress in HEP, astrophysics, and cosmology is in improving sensitivities and bounds rather than discovering hitherto unknown patterns in data or new physics.
It seems clear, however, that interpretable and explainable ML will play a significant role in explaining discoveries yet to come, as the line between ML and classical physics models continues to blur, and thus it is best to be equipped in the present with the tools to understand what exactly it is that our models are telling us.

\section*{Acknowledgements}

We thank the organizers and members of VERaiPHY for organizing this review and many fruitful discussions.
In particular, RG thanks Lydia Brenner, Lukas Heinrich, Gaia Grosso, Mikael Kuusela, Noam Levi, and Louis Lyons for discussions and useful feedback on interpretability and explainability.
LLS thanks Hiranya Peiris and Andrew Pontzen for insightful discussions on the topics of interpretability and explainability over the past years.
We also thank Hiranya Peiris for useful comments that improved this manuscript.

% TODO: include author contributions
\paragraph{Author contributions}
RG and LLS were equal contributors and main authors to all aspects of this review article.  JT provided advice about article scope and high-level content, and performed light editing.

% TODO: include funding information
\paragraph{Funding information}
LLS acknowledges funding by the Deutsche Forschungsgemeinschaft (DFG, German Research Foundation) under Germany’s Excellence Strategy -- EXC 2121 ``Quantum Universe'' -- 390833306 and via the SciFM consortium (05D25GU4) funded by the German Federal Ministry of Research, Technology, and Space (BMFTR) in the ErUM-Data action plan.
RG acknowledges support in part by DOE grants
DE-SC0026301 and DE-SC101977, and by NSF grants OAC-2103889, OAC-2411215, and OAC-2417682.
JT was supported by the National Science Foundation under Cooperative Agreement
PHY-2019786 (The NSF Institute for Artificial Intelligence and Fundamental Interactions, 
\url{http://iaifi.org/}), by the U.S.\ Department of Energy Office of High Energy Physics under grant
number DE-SC0012567, by the Simons Foundation through Investigator grant 929241, and he thanks the Institut des Hautes \'Etudes Scientifiques (IHES) and the Institut de Physique Th\'eorique (IPhT) for providing an inspiring sabbatical environment.

%Authors are required to provide funding information, including relevant agencies and grant numbers with linked author's initials. Correctly-provided data will be linked to funders listed in the \href{https://www.crossref.org/services/funder-registry/}{\sf Fundref registry}.


% Provide your bibliography here. You have two options:

% Use your bibtex library, formatted by the SciPost style file.
% \bibliographystyle{SciPost_bibstyle}
\bibliography{refs.bib}

%%%%%%%%%% END TODO: BIBLIOGRAPHY

\end{document}